# Experimental Study of Through-the-Wall Respiration Sign Detection Using Ultra-wideband Impulse Radar

Budiman P.A. Rohman[1,2,*], Muhammad Bagus Andra[1], Dion Hayu Fandiantoro[1], Masahiko Nishimoto[1]
[1]Graduate School of Science and Technology, Kumamoto University, Japan
[2]Indonesian Institute of Sciences, Indonesia
budi028@lipi.go.id, nisimoto@cs.kumamoto-u.ac.jp

*Abstract*- The through-the-wall human being detection and localization in the complex environments by using the radar system has many possible applications such as for law enforcement, and search and rescue missions in disaster-stricken areas. This paper discusses the experimental study of ultra-wideband impulse radar for obscured human respiration detection through spectrum analysis. We investigate the effect of the range of human-wall and radar-wall to the peak factor value of human respiration sign power spectrum. This result will be useful as a reference for analyzing the possibility of the sign to be detected by the radar with respect to the distance from both sides. According to the experimental results, longer both distance increases gradually the peak factor value. However, the peak became narrower. Besides, at a very short distance between areas and walls, the peak factor became very small because of the strong reflection of both objects. Therefore, an adaptive threshold technique and background suppression are needed in order to maintain the detection performance against any possible conditions.

*Index Terms*- through-the-wall, ultra-wideband impulse radar, buried survivor, human vital sign, disaster.

## I. INTRODUCTION

THE uses radar for detecting the through-the-wall human vital sign became an interest of many researchers. There are many possible applications of this system such as for security and disaster search and rescue mission [1-3]. This system is possible to be applied in the moving platform such as an unmanned aerial vehicle or mobile robot for a large observation mission. The use of these moving and compact platforms is promising to be conducted especially in the disaster-stricken area to search the buried survivor [1].

Various signal processing technique regarding this case has been proposed by several researchers [2-5]. However, those proposed approaches have a complex algorithm that will require a huge computation resource and time. Thus, those proposed techniques are not recommended to be implemented in the moving platform such as a small-sized drone which has many limitations including the low-performance computation and limited power resource.

Considering this issue, we conduct an experimental study about the use of ultra-wideband (UWB) impulse radar for detecting the through-the-wall human respiration sign by implementing the primary vital sign spectrum processing technique. For evaluation, we use a single layer brick wall with various distances of both between human-wall and radar-wall. Through the peak factor value analysis on the frequency-limited of spectrum signal, we got the information about the possibility of radar to detect the human respiration by applying this approach.

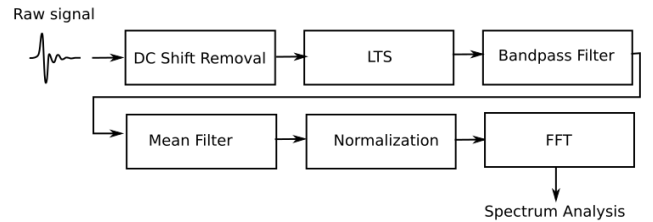

Fig. 1. Primary UWB radar signal processing for vital sign detection

## II. RADAR SIGNAL PROCESSING

Assuming that $X \in R^{M \times N}$ is the two dimensional signal recorded by radar in a certain duration where *M* and *N* corresponding with the fast time and slow time domain, respectively. Simply, the fast time-domain reflects the object distance while the slow time domain brings the respiration information including its periodic pattern.

The signal processing in this study implements primary techniques commonly used in obscured vital signal detection researches [3]. The step of signal processing is shown in Fig.1. The signal recorded from the UWB radar is pre-processed using DC-shift removal and linear trend subtraction in slow time domain. Then, the signal is filtered by the band-pass filter, mean filter with window size five samples which are then normalized. After that, each signal in the slow time domain is converted to the frequency domain by Fast Fourier Transform to extract the human vital sign frequency. Since the heartbeat signal is very weak, in this study, we only consider the respiration signal with frequency range 0.3-0.8 Hz. Thus, we make a window for limiting around that frequency and calculate the peak factor of the corresponding respiration sign spectrum. Assuming that the radar signal in the frequency domain is *Y* so that the peak factor value is calculated as follows,

$$P = \frac{Y_{max}}{Y_{rms}} \qquad (1)$$

where $Y_{max}$ is the peak value corresponding to the respiration signal and $Y_{rms}$ is the root-mean-squared value of windowed spectrum.

## III. EXPERIMENT AND ANALYSIS

In this study, we use UWB impulse radar which transmits a first-order differentiated Gaussian pulse with bandwidth 0.9GHz-6.5GHz. The sampling rate of raw data obtained is around 39GS/s while each of the data consists of 200 traces where each trace has 1024 sample points. The experiment setup for taking data is shown in Fig. 2.

From the results showed in Table 1, it can be seen obviously that in the observed environment, the distance has an effect on the peak factor of the vital sign. Simply, when the range increases, the peak factor became higher and the peak became sharper. It will make easy the detection process but if the peak is too narrow, it may be masked by noise. Of course, at a certain distance, human respiration signature may be invisible because of the electromagnetic wave attenuation. These results are confirmed by Fig. 3.

## IV. CONCLUSIONS

This paper discussed the experimental study about the application of UWB impulse radar for detection the human respiration sign from the back of the wall. Through the peak factor analysis, the detection can be done by implementing primary signal processing steps. The experimental results reveal that the peak factor of human respiration signal increases following the range increase. However, the peak shape became narrower. This tendency may change at a certain distance because of the wave attenuation. Other possible conditions should be considered to evaluate more the detectability of the human respiration sign behind wall by applying a primary radar signal processing technique.

## V. ACKNOWLEDGMENT

This research is supported in part by IEEE Geosciences and Remote Sensing Society through IEEE-GRSS GSC 2018.

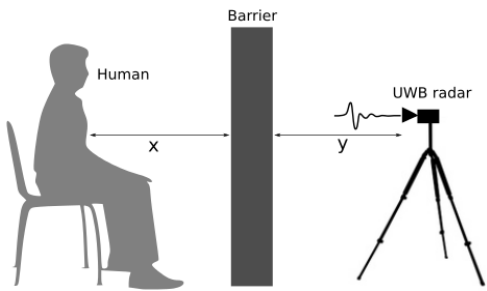

Fig. 2. Experiment setup

Table 1. Comparison of Peak Factor with various distance

| Case | Range (cm) | | Peak Factor |
|---|---|---|---|
| | Body-Wall (x) | Wall-Radar (y) | |
| 1 | 40 | 40 | 6.87 |
| 2 | | 160 | 7.35 |
| 3 | 80 | 40 | 7.39 |
| 4 | | 160 | 9.79 |

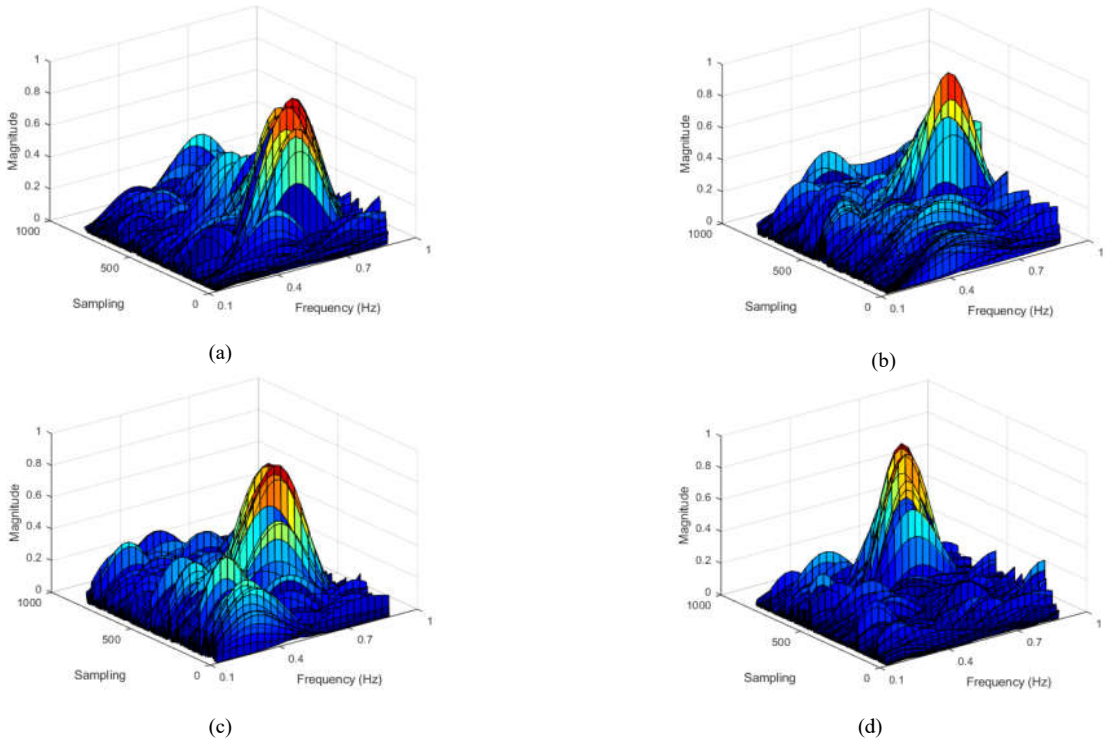

(a)    (b)

(c)    (d)

Fig. 3. Result of signal processing with range value x/y : (a) 40/40, (b) 40/160, (c) 80/40, and (d) 80/160